\documentclass[conference]{IEEEtran}
\IEEEoverridecommandlockouts

\usepackage{cite}
\usepackage{amsmath,amssymb,amsfonts}
\usepackage{algorithmic}
\usepackage{graphicx}
\usepackage{textcomp}
\usepackage{xcolor}
\usepackage{xspace,enumitem,booktabs}

\def\BibTeX{{\rm B\kern-.05em{\sc i\kern-.025em b}\kern-.08em
    T\kern-.1667em\lower.7ex\hbox{E}\kern-.125emX}}
\begin{document}

\title{Investigating the Effects of Diffusion-based Conditional Generative Speech Models Used for Speech Enhancement on Dysarthric Speech}

\author{

\IEEEauthorblockN{1\textsuperscript{st} Joanna Reszka}
\IEEEauthorblockA{\textit{CVSA, Bayer AG}\\
Warsaw, Poland \\
joanna.reszka@bayer.com}

\and
\IEEEauthorblockN{2\textsuperscript{nd} Parvaneh Janbakhshi}
\IEEEauthorblockA{\textit{CVSA, Bayer AG}\\
Berlin, Germany \\
parvaneh.janbakhshi@bayer.com}
\and
\IEEEauthorblockN{3\textsuperscript{rd} Tilak Purohit}
\IEEEauthorblockA{\textit{Idiap Research Institute}\\
Martigny, Switzerland \\
tilak.purohit@idiap.ch }
\and
\IEEEauthorblockN{4\textsuperscript{th} Sadegh Mohammadi}
\IEEEauthorblockA{\textit{CVSA, Bayer AG} \\
Berlin, Germany \\
sadegh.mohammadi@bayer.com }
}

\maketitle
\begin{abstract}
In this study, we aim to explore the effect of pre-trained conditional generative speech models for the first time on dysarthric speech due to Parkinson's disease recorded in an ideal/non-noisy condition. Considering one category of generative models, i.e., diffusion-based speech enhancement, these models are previously trained to learn the distribution of clean (i.e, recorded in a noise-free environment) typical speech signals. Therefore, we hypothesized that when being exposed to dysarthric speech they might remove the unseen atypical paralinguistic cues during the enhancement process. By considering the automatic dysarthric speech detection task, in this study, we experimentally show that during the enhancement process of dysarthric speech data recorded in an ideal non-noisy environment, some of the acoustic dysarthric speech cues are lost. Therefore such pre-trained models are not yet suitable in the context of dysarthric speech enhancement since they manipulate the pathological speech cues when they process clean dysarthric speech. Furthermore, we show that the removed acoustics cues by the enhancement models in the form of residue speech signal can provide complementary dysarthric cues when fused with the original input speech signal in the feature space.

\end{abstract}


\section{Introduction}

Dysarthria is a motor speech disorder arising from impairments in speech production mechanisms. 
Neurodegenerative disorders such as Parkinson’s disease (PD)
can cause speech dysarthria which without further PD management will lead to reduced communicative ability and life quality~\cite{Darley1969}. For the past two decades, researchers focused on developing automatic machine learning techniques especially aiming at dysarthric speech detection, i.e., discriminating between speech from healthy and dysarthric (e.g., PD) patients. Such technologies offer reliable, objective, and cost-effective speech assessments to aid clinicians in their inherent subjective and time-consuming auditory-perceptual analyses~\cite{Baghai-Ravary2012}

The majority of techniques in literature for dysarthric speech assessment deal with handcrafting acoustic features followed by classical machine learning analysis~\cite{gillespie17_interspeech, np18_interspeech, Kodrasi2020, Lahoti2022, Suppa2022,kovac_exploring_2024}.
To learn more complicated but discriminative speech representations for such a task, deep learning approaches have been also used.
One category of deep learning approaches involves typical end-to-end training networks for the primary task of dysarthric speech detection, e.g., convolutional neural networks (CNNs)~\cite{vasquezcorrea17_interspeech, Kodrasi2021, GUPTA2021105, Janbakhshi2021, Janbakhshi2022}.
In deep learning approaches many strategies have been previously proposed to learn more robust features and alleviate overfitting problems associated with the availability of a small amount of pathological speech training data:
i) modifying training strategies such as incorporating pairwise training as in~\cite{Bhati2019, Janbakhshi2021} 
ii) multi-task learning in which supervised or unsupervised training with an additional secondary but relevant task on pathological speech data is considered~\cite{vasquezcorrea18_interspeech, Janbakhshi2021b, janbakhshi22_interspeech}, iii) transfer learning, in which pre-trained networks have been used to extract features from dysarthric speech where these models are initially trained using larger datasets for different tasks, e.g., speech reconstruction as in auto-encoders~\cite{VASQUEZCORREA2020, MILLER2024}, supervised models as in speech attributes or phonological features extractor networks~\cite{Janbakhshi2021, Liu2023}, and also self-supervised models, e.g., wav2vec2 as used in~\cite{javanmardi_pre-trained_2024, Schu2023,laquatra24_interspeech} for dysarthric speech detection.

In the context of speech generation, generative models to produce realistic speech samples have gained immense interest.
Such models capture the underlying patterns or distribution from a given speech training data. Since the potential and effects of speech generative models on dysarthric speech is under-explored, here we aim to investigate the effect of such models on dysarthric speech.

In this work, we focused on one category of generative models, i.e., conditional generative models for speech enhancement in which the aim is to recover clean speech signals given the noisy audio recordings. These generative models unlike their discriminative counterparts aim to learn a prior distribution over clean, typical speech data based on the inherent speech spectral and temporal patterns~\cite{Richter2022SpeechEA}. These models are trained to produce clean, typical/healthy speech samples (dictated by the properties of their seen training data that most likely do not include atypical pathological speech). We hypothesize that when instead of noisy speech signals these models are exposed to clean dysarthric speech signals, during their enhancement process, they may remove some of the local acoustic pathological cues.
Verification of such a hypothesis can be a step towards shedding light on the following: i) in the context of dysarthric speech enhancement where the goal is to preserve dysarthric cues while removing other unwanted speech noise patterns (e.g., background noise), will pre-trained enhancement models suitable, i.e., will they preserve acoustic cues of dysarthric speech even if signals are non-noisy/clean? ii) in the context of dysarthric speech assessment, can the residue signals (what have been removed from original speech during the process) from such models provide complementary information about dysarthric speech? In this work, we do not aim to enhance noisy dysarthric speech using a generative speech enhancement model as in~\cite{laquatra24_interspeech} since our considered dataset is free from noise (cf. \ref{data}). Our goal is to investigate how conditional generative speech enhancement as one category of the generative speech model reacts to dysarthric (clean) speech.

It is worth mentioning that the dysarthric speech detection approach proposed by~\cite{VASQUEZCORREA2020, MILLER2024} can be also seen as another view of our hypothesis but in the signal reconstruction paradigm using auto-encoders. In~\cite{VASQUEZCORREA2020, MILLER2024}, an auto-encoder pre-trained on a larger healthy speech dataset is used to extract and compute dysarthric speech reconstruction error (due presence of paralinguistic features) which served as a novel feature set for dysarthric speech detection. However, in this paper, we aim to verify our hypothesis by investigating the effects of pre-trained conditional generative models meant to learn the distribution of typical clean speech on dysarthric speech.

Due to the recent advancement of diffusion-based generative models for speech enhancement, in this work, we focused on investigating the effects of three pre-trained diffusion-based models on dysarthric speech. We first experimentally investigate how the processed dysarthric speech as the output of diffusion-based speech enhancement models, can affect the performance of dysarthric speech detection task using different classification methods. Then, we experimentally verify if the residue signal, from speech enhancement models can provide complementary information compared to the original speech signals benefiting the performance of the dysarthric speech detection task.
Section~\ref{DM} introduces the considered pre-trained diffusion-based speech enhancement models along with the methods considered for dysarthric speech detection task. Section~\ref{expsetup} presents the considered dataset and experimental setup, Section~\ref{results} presents the experimental results and finally, Section~\ref{conclusion} concludes the paper.

\section{Diffusion-based speech enhancement impacts on dysarthric speech}\label{DM}
This section introduces the three considered diffusion-based generative models used for speech enhancement. Furthermore, the dysarthric speech analyses are presented.

\begin{figure}[b!]
  \centering
  \vspace{-5pt}
  \includegraphics[width=\linewidth]{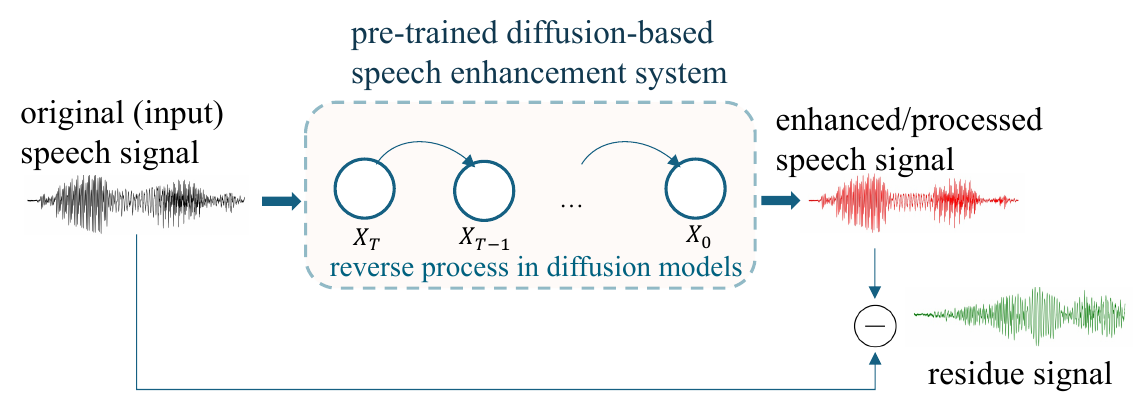}\vspace{-2pt}
  \caption{Schematic representation of speech enhancement models to obtain enhanced and residue signals.}
  \label{fig:block_diag}
    \vspace{-1pt}
\end{figure}

\subsection{Diffusion-based speech enhancement models}
We considered three pre-trained diffusion-based speech enhancement models introduced in the following.
\subsubsection{Score-based generative model for speech enhancement (SGMSE)}
SGMSE is a diffusion process based on a stochastic differential equation for speech enhancement that operates on the complex short-time Fourier transform domain. As opposed to usual generation tasks, instead of a reverse process from pure Gaussian noise, in SGMSE, a mixture of noisy speech and Gaussian noise is used. SGMSE is reported to compete with recent discriminative speech enhancement models and achieve better generalization when evaluating cross-dataset scenarios with unseen noise sources during training. It has been also shown that the SGMSE framework can be used to train individual models for different distortion types, e.g., additive noise and non-additive corruptions such as reverberation~\cite{Richter2022SpeechEA}.

\subsubsection{Conditional diffusion probabilistic model for speech enhancement (CDiffuSE)}
CDiffuSE is a generalized formulation of the diffusion probabilistic model for speech enhancement that in its reverse process, can adapt to non-Gaussian real noises. Such noise characteristics are incorporated by linearly interpolating between clean and noisy speech along the diffusion process. CDiffuSE model performs enhancement in the time domain.
This model has shown generalization capabilities to speech data with unseen noise characteristics~\cite{Lu2022}.

\subsubsection{DiffWave}
DiffWave is a diffusion probabilistic model for conditional and unconditional waveform generation. It has been shown that the unconditional DiffWave model can perform zero-shot speech denoising without knowing the added noise types, since DiffWave learns a good prior of typical raw audio waveform~\cite{kong2021diffwave}.

\subsection{Enhanced/processed and residue signals}
Upon applying each of the pre-trained generative speech enhancement models to our original speech signals from either healthy or dysarthric speakers, supposedly enhanced or processed signals will be obtained. Furthermore, by subtracting the original input speech signal from its corresponding processed signal, a residue signal will be computed. Using a residue signal is inspired from~\cite{VASQUEZCORREA2020, MILLER2024}.
In practice, we have observed that residue signals have speech characteristics. By analyzing both processed and residue signals obtained from each enhancement model, we aim to verify whether some of the dysarthric acoustic cues will be lost during the enhancement process and therefore will be added to the residue signals. Figure~\ref{fig:block_diag} depicts a schematic representation of diffusion-based enhancement models to obtain processed and residue signals.

\subsection{Dysarthric processed and residue speech signals analysis}
\begin{figure}[t!]
  \centering
   \vspace{-5pt}
  \includegraphics[width=\linewidth]{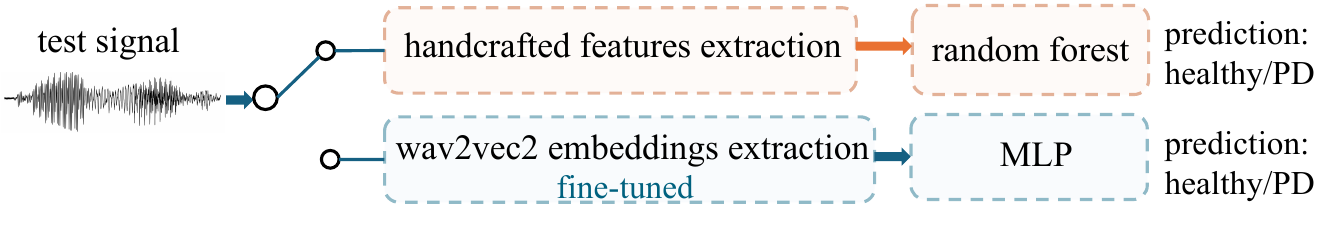}\vspace{-2pt}
  \caption{Schematic representation of dysarthric (e.g., PD) speech detection.}
  \label{fig:block_diag2}
   \vspace{-1pt}
\end{figure}

To gain insights into the effects of generative speech enhancement models on dysarthric speech, processed and residue signals produced by these models have been used for automatic dysarthric speech detection task. For this purpose, two state-of-the-art dysarthric speech detection approaches, one based on deep learning and one based on machine learning using handcrafted features, are considered.
Figure~\ref{fig:block_diag2} depicts the schematic representation of the considered dysarthric speech detection approaches.

\subsubsection{Wav2vec2$+$MLP}
Wav2vec2 is a state-of-the-art self-supervised learning model operating directly on raw waveforms to produce robust latent speech representations that have shown impressive results on many downstream speech tasks~\cite{yang21c_interspeech}. Analyzing processed and residue speech from dysarthric speakers in comparison to original speech signals used to pretrain the model can result in a domain shift. Therefore, dysarthric speech detection is achieved by fine-tuning the wav2vec2 CNN encoder model with a lightweight multilayer perceptron (MLP) classifier. To assess the fusion of two types of input signals at feature level (cf.~\ref{results}), the two feature sets produced by the wav2vec2 encoder are fused before the classifier using weighted attention pooling~\cite{Truong2024} inspired from squeeze-and-excitation block in~\cite{Jie2018}.

\subsubsection{OpenSMILE$+$RF}
Brute-force Handcrafted OpenSMILE feature set from the ComParE-2016 challenge~\cite{schuller16_interspeech} provided by OpenSMILE toolkit~\cite{Eyben2010}
has served as a baseline feature set for many speech paralinguistic tasks~\cite{Eyben2010, schuller16_interspeech}. In this work, dysarthric speech detection is also achieved by training a random forest (RF) classifier on OpenSMILE features set. To assess the fusion of two types of input signals (cf.~\ref{results}), OpenSMILE feature sets of both signals are concatenated before being fed to the classifier.

\section{Experimental Setup}\label{expsetup}
In this section, the considered dysarthric speech dataset, experimental setup, and evaluation protocols are presented.
\subsection{Dataset} \label{data}
For the dysarthric speech detection task, we considered Spanish recordings from $50$ PD patients ($25$ males,
$25$ females) and $50$ healthy speakers ($25$ males, $25$ females) from the PC-GITA database~\cite{Orozco2014}. PC-GITA speech dataset is recorded in a soundproof booth using professional audio equipment ensuring an optimal environment free from noise. For each speaker recordings of utterances (with the same phonetic contents across speakers), i.e., $24$ words, $10$ sentences, and $1$ text recorded initially at a sampling frequency of $44.1$ kHz are used.
For our analysis, all recordings first have been downsampled to $16$ kHz. Silence segments from words are manually removed while for other recordings an energy-based voice activity detector~\cite{Boersma2001} has been used.

\subsection{Training and evaluation}

For the dysarthric speech detection task, we resort to a stratified 10-fold speaker-independent cross-validation evaluation quantified by metrics: accuracy, area under ROC curve (AUC), sensitivity (correct classification rate of patients), and specificity (correct classification rate of healthy speakers). The final performance is the mean and standard deviation of the speaker-level classification metric values obtained across $10$ folds.

For training the wav2vec2$+$MLP, we preprocessed the speech data by first concatenating all speech modules for each speaker and then segmenting the resulting utterance into chunks of 3-second segments. To monitor the training process in each fold, a validation set excluded from the training fold containing data from $10$ speakers is used.
The MLP trained on wav2vec2 embeddings consists of two
fully connected layers with $256$ hidden units in between. Similar to~\cite{yang21c_interspeech}, frame-level wav2vec2 embeddings (of 
 dimension {$768$}) are mean-pooled across each input utterance before being fed into the MLP classifier. wav2vec2$+$MLP is trained with a batch size of $32$ for $30$ epoch supervised by cross-entropy loss. While the MLP classifier is trained with a learning rate of $0.001$, the wav2vec2 encoder is fine-tuned with a $10$ times smaller learning rate to alleviate over-fitting issues. To compute speaker-level classification performance, segment-level predictions through soft voting for each speaker are aggregated.

For training the RF on the OpenSMILE feature set, in each training fold, 
a nested cross-validation strategy with an internal $9$
folds for hyper-parameter optimization based on the accuracy
of the training set is used.

Regarding the speech enhancement models, in this paper, their released pre-trained models are used (the reader is referred to~\cite{Richter2022SpeechEA,Lu2022,kong2021diffwave} for details on the models and their corresponding training data).

\section{Experimental Results}\label{results}



\begin{table}[th]
\vspace{-1pt}
  \caption{Performance of dysarthric speech detection using wav2vec2 encoder$+$MLP processing original, processed and residue speech signals produced by different speech enhancement models.}\vspace{-1pt}
  \label{tab1}
  \centering
  \resizebox{\columnwidth}{!}{
\begin{tabular}{lllll}
\toprule
\multicolumn{1}{l}{input signal}                                       & \multicolumn{1}{c}{AUC} & \multicolumn{1}{c}{accuracy} & \multicolumn{1}{c}{sensitivity} & \multicolumn{1}{c}{specificity} \\\midrule[0.3pt]\midrule[0.3pt]
original                                               &  $0.83 \pm 0.11$                      &       $79 \pm 10$                       &        $74 \pm 20$                       &     $84 \pm 17$                       
\\\midrule[0.3pt]
\multicolumn{5}{c}{SGMSE}    \\
\midrule[0.3pt]
processed                                               &   $0.79 \pm 0.16$                      &       $69 \pm 14$                       &             $68 \pm 20$                    &     ${70} \pm 20$                            \\
residue &                           $0.88 \pm 0.12$                           &        $79 \pm 14$                         &    $84 \pm 20$         & $74 \pm 16$                    \\
residue\&original&                         $0.91 \pm 0.09$                             &    $82 \pm 13$                             &   $82 \pm 17$            & $82 \pm 23$                \\ 
\midrule[0.3pt]
\multicolumn{5}{c}{CDiffuSE}                                                                                                                                                            \\
\midrule[0.3pt]
processed                                               &  ${0.85} \pm 0.09$ & $74 \pm 12$    & $70 \pm 18$ & $78 \pm 22$ \\
residue &      ${ 0.89} \pm 0.07$                   &     ${ 81} \pm 9$                         &         ${ 80} \pm 15$                        &         $ 82 \pm 21$                        \\
residue\&original&                         $0.88 \pm 0.08$                             &    $78 \pm 7$                             &   $74 \pm 20$            & $82 \pm 19$                \\ 
\midrule[0.3pt]
\multicolumn{5}{c}{DiffWave}\\
\midrule[0.3pt]
processed                                               & $0.74 \pm 0.10$                         &      $68 \pm 12$                        &       $70 \pm 13$                          &    $66 \pm 25$                             \\
residue&                         $0.82 \pm 0.11$                             &    $79 \pm 9$                             &   $82 \pm 14$            & $76 \pm 25$                \\ 
residue\&original&                         $0.86 \pm 0.09$                             &    $82 \pm 9$                             &   $76 \pm 20$            & $88 \pm 13$                \\ 
\bottomrule
\end{tabular}
}
\end{table}

\begin{table}[th]
\vspace{-1pt}
  \caption{Performance of dysarthric speech detection using OpenSMILE$+$RF processing original, processed and residue speech signals produced by different speech enhancement models.}\vspace{-1pt}
  \label{tab2}
  \centering
  \resizebox{\columnwidth}{!}{
\begin{tabular}{lllll}
\toprule
\multicolumn{1}{l}{input signal}                                       & \multicolumn{1}{c}{AUC} & \multicolumn{1}{c}{accuracy} & \multicolumn{1}{c}{sensitivity} & \multicolumn{1}{c}{specificity} \\\midrule[0.3pt]\midrule[0.3pt]
original                                               &  $0.82 \pm 0.10$                      &       $74 \pm 11$                       &        $76 \pm 17$                       &     $72 \pm 18$                       
\\\midrule[0.3pt]
\multicolumn{5}{c}{SGMSE}    \\
\midrule[0.3pt]
processed                                               &   $0.82 \pm 0.13$                      &       $74 \pm 11$                       &             $66 \pm 22$                    &     ${82} \pm 19$                            \\
residue &                           $0.68 \pm 0.18$                           &        $66 \pm 12$                         &    $60 \pm 22$         & $71 \pm 16$                    \\
residue\&original&                         $0.85 \pm 0.11$                             &    $75 \pm 9$                             &   $76 \pm 20$            & $74 \pm 18$                \\ 
\midrule[0.3pt]
\multicolumn{5}{c}{CDiffuSE}                                                                                                                                                            \\
\midrule[0.3pt]
processed                                               &  ${0.83} \pm 0.08$ & $74 \pm 9$    & $74 \pm 16$ & $74 \pm 16$ \\
residue &      ${0.87} \pm 0.11$                   &     ${79} \pm 7$                         &         ${78} \pm 17$                        &         $ 80 \pm 15$                        \\
residue\&original&                         $0.84 \pm 0.11$                             &    $78 \pm 10$                             &   $82 \pm 19$            & $74 \pm 16$                \\ 
\midrule[0.3pt]
\multicolumn{5}{c}{DiffWave}\\
\midrule[0.3pt]
processed                                               & $0.73 \pm 0.13$                         &      $69 \pm 11$                        &       $66 \pm 18$                          &    $72 \pm 16$                             \\
residue&                         $0.86 \pm 0.10$                             &    $75 \pm 11$                             &   $76 \pm 20$            & $74 \pm 18$                \\ 
residue\&original&                         $0.84 \pm 0.11$                             &    $78 \pm 6$                             &   $80 \pm 13$            & $76 \pm 17$                \\ 
\bottomrule
\end{tabular}
}
\end{table}

In this section, the performance of dysarthric speech detection on speech signals from the considered dysarthric dataset is evaluated and compared to their corresponding processed and residue speech signals after being processed by different diffusion-based generative speech enhancement models, i.e., SGMSE, CDiffuSE, and DiffWave.


Table~\ref{tab1} and Table~\ref{tab2} presents the performance of dysarthric speech detection task using wav2vec2$+$ MLP and OpenSMILE$+$RF processing original speech signal, as well as processed and residue speech signals produced by different speech enhancement models. Considering both dysarthric speech detection methods, the following observations can be made. First, the overall performance of dysarthric speech detection using processed signals from all three enhancement models is comparable or lower than using original signals but with sensitivity values (correct classification rate of patients) being consistently lower than using original signals. This is aligned with our hypothesis that such enhancement models can remove some dysarthric speech cues which translates into lower sensitivity obtained using processed signals.
Second, we observe that the residue signals from CDiffuSE and DiffWave in both dysarthric speech detection methods achieved overall better performance than their corresponding processed signals showing they contain some of the dysarthric cues lost in processed signals which even boosted the performance. Furthermore, the better or comparable performance obtained using the residue signals as opposed to the original signals in both detection methods suggests that these signals still contain some dysarthric speech cues.
Third, after fusing original and residue signals at the feature level for each of the models, an overall performance boost compared to the original signals (except a comparable performance achieved by only CDiffuse using wav2vec2$+$ MLP) is achieved which can suggest that residue signals contain useful dysarthric speech cues that can act as complementary information for dysarthric speech detection task.

It should be noted that the large standard deviations arise due to the small size of test folds ($5$ PD $+$ $5$ healthy speakers). Therefore, changes in the prediction of one speaker in each test fold contribute to $10\%$ and $20\%$ changes in accuracy and sensitivity or specificity metrics, respectively.

In summary, the presented results show that pre-trained diffusion-based speech enhancement models that are meant to learn clean typical speech distribution during the enhancement process can remove acoustic cues necessary for dysarthric speech detection. This suggests such out-of-box models are not yet suitable for speech enhancement of pathological speech. Furthermore, residue signals containing information lost during the enhancement process, can provide complementary dysarthric speech cues. This can be a preliminary step towards validating the hypothesis that the generative models aim to model characteristics of clean typical speech can react to dysarthric speech characteristics as a deviation compared to their learned model. In the future when such models are more powerful in modeling diverse clean typical speech and in adapting to diverse noise sources, it can be expected that they can be more sensitive in reacting to atypical paralinguistic cues present in speech.

\section{Conclusion}\label{conclusion}
In this study, for the first time, we explored the effect of pre-trained diffusion-based conditional generative models used for speech enhancement on dysarthric speech due to PD. These models, instead of noisy speech signals, received clean dysarthric speech signals as input. This study was motivated by the hypothesis that pre-trained generative speech enhancement approaches modeling clean typical speech distribution can react to atypical paralinguistic cues in clean dysarthric speech. We stepped toward verifying this hypothesis by performing an automatic dysarthric speech detection task on processed signals and residue signals using three conditional speech generative speech enhancement systems. Our experimental results suggested that during the enhancement process by different models, some dysarthric speech cues are lost since it resulted in a lower dysarthric speech detection performance. Furthermore, such removed acoustics cues by the models in the form of residue signal provided complementary dysarthric speech information to the original input signal which further improved the performance of automatic dysarthric speech detection.

\bibliographystyle{IEEEtran}
\footnotesize

\end{document}